\documentclass[preprint,showpacs,preprintnumbers,amsmath,amssymb,prl]{revtex4}
\usepackage[pdftex]{graphicx}
\usepackage{bm}
\usepackage{float}
\usepackage{color}
\usepackage{hyperref}
\usepackage{pdfpages}
\usepackage[T1]{fontenc}
\begin{document}
\title{\color{blue}Kinetics of the glass transition of fragile soft colloidal suspensions}
\author{Debasish Saha}
\email{debasish@rri.res.in}
\affiliation{Soft Condensed Matter Group, Raman Research Institute, C. V. Raman Avenue, Sadashivanagar, Bangalore 560 080, INDIA}
\author{Yogesh M Joshi}
\email{joshi@iitk.ac.in}
\affiliation{Department of Chemical Engineering, Indian Institute of Technology Kanpur, Kanpur 208 016, INDIA.}
\author{Ranjini Bandyopadhyay}
\email{ranjini@rri.res.in}
\affiliation{Soft Condensed Matter Group, Raman Research Institute, C. V. Raman Avenue, Sadashivanagar, Bangalore 560 080, INDIA}
\vspace{0.5cm}
\date{\today}
\begin{abstract}
Microscopic relaxation timescales are estimated from the autocorrelation functions obtained by dynamic light scattering experiments for Laponite suspensions with different concentrations ($C_{L}$), added salt concentrations ($C_{S}$) and temperatures ($T$). It has been shown in an earlier work [Soft Matter {\bf 10}, 3292 (2014)] that the evolutions of relaxation timescales of colloidal glasses can be compared with molecular glass formers by mapping the waiting time ($t_{w}$) of the former with the inverse of thermodynamic temperature ($1/T$) of the latter. In this work, the fragility parameter $D$, which signifies the deviation from Arrhenius behavior, is obtained from fits to the time evolutions of the structural relaxation timescales. For the Laponite suspensions studied in this work, $D$ is seen to be independent of $C_{L}$ and $C_{S}$, but is weakly dependent on $T$. Interestingly, the behavior of $D$ corroborates the behavior of fragility in molecular glass formers with respect to equivalent variables. Furthermore, the stretching exponent $\beta$, which quantifies the width $w$ of the spectrum of structural relaxation timescales is seen to depend on $t_{w}$. A hypothetical Kauzmann time $t_{k}$, analogous to the Kauzmann temperature for molecular glasses, is defined as the timescale at which $w$ diverges. Corresponding to the Vogel temperature defined for molecular glasses, a hypothetical Vogel time $t^{\infty}_{\alpha}$ is also defined as the time at which the structural relaxation time diverges. Interestingly, a correlation is observed between $t_{k}$ and $t^{\infty}_{\alpha}$, which is remarkably similar to that known for fragile molecular glass formers. A coupling model that accounts for the $t_{w}$-dependence of the stretching exponent is used to analyse and explain the observed correlation between $t_{k}$ and $t^{\infty}_{\alpha}$.
\end{abstract}
\maketitle
\section{Introduction}
The dependence of transport properties (diffusivity, viscosity etc.) and structural relaxation times near the glass transition are of crucial importance in understanding glass formers. A few key observations, like the rapid increase of viscosity near the glass transition temperature $T_{g}$, the heat-capacity jump at $T_{g}$, the Kauzmann entropy catastrophe, the super-Arrhenius temperature dependence of the structural relaxation processes, and fragile behavior are common to many glass formers \cite{Angell_Science_1995_IGS,Lubchenko_Wolynes_Annu_Rev_Phys_Chem_2007_IGS,Angell_J_Res_Natl_Inst_Stand_Technol_1997_IGS}. An enormous increase in viscosity and relaxation time (14 decades and more) is observed as a molecular glass former is quenched towards its glass transition temperature \cite{Ediger_Angell_NageL_JPC_1996_IGS}. Specific heat measurements show a jump in the heat-capacity at $T_{g}$ and the extent of the jump, in general, is larger for a fragile glass \cite{Lubchenko_Wolynes_Annu_Rev_Phys_Chem_2007_IGS}. Fragile glasses show super-Arrhenius temperature dependence and exhibit an extremely steep increase in viscosity $\eta$ which can be expressed by the Vogel-Fulcher-Tammann (VFT) relation, i.e., $\eta=\eta_{0}\exp\left[DT_{0}/(T-T_{0})\right]$. Here, $T_{0}$ is the Vogel temperature at which $\eta$ diverges. The fragility parameter $D$ quantifies the deviation from Arrhenius behavior. $D$ is a material specific quantity \cite{Angell_J_Non_Cryst_Solids_1991_IGS}, whose magnitude is small (typically $<$10) for fragile supercooled liquids and can change to very large values for strong glass formers. It is very difficult to differentiate between an Arrhenius temperature dependence and a super-Arrhenius temperature dependence for $D>100$ \cite{Bohmer_Ngai_Angell_Plazek_JCP_1993_IGS,Angell_J_Non_Cryst_Solids_1991_IGS}.\\
\indent The relation between the fragility parameter $D$ and the structural properties of a material is still not completely understood despite many theoretical and experimental studies. A correlation is drawn between the fragility of a material and its physical properties, i.e. its Poisson's ratio or the relative strength of its shear and bulk moduli \cite{Novikov_Sokolov_Nature_2004_IGS}. The relation between the nature of the interaction potential and fragility has also been studied for model binary mixture glass formers \cite{Sengupta_Sastry_JCM_2011_IGS} and colloidal glass formers \cite{Mattsson_Miyazaki_Nature_2009_IGS}. In the supercooled liquids literature, $T_{g}$ is the temperature at which the mean $\alpha$-relaxation time ($<\tau_{\alpha}>$) is 100 sec \cite{Bohmer_Ngai_Angell_Plazek_JCP_1993_IGS,Angell_J_Non_Cryst_Solids_1991_IGS}. It is to be noted that $T_{g}$ depends on the heating or cooling rate \cite{Moynihan_et_al_JPC_1974_IGS}.\\
\indent The phenomenology of glass formers have often been described by a potential energy landscape (PEL) \cite{Debenedetti_Stillinger_Nature_2001_IGS,Heuer_Topical_review_PEL_JPCM_2008_IGS} which can be visualized in terms of a multi-dimensional surface describing the dependence of the potential energy of the system as a function of the particle coordinates \cite{Stillinger_Science_1995_IGS}. If $N$ is the number of particles, then the dimension of the hyperspace is $3N+1$, with the system represented by a point evolving temporally in this complex potential hyperspace \cite{Stillinger_Science_1995_IGS}. In this topographic description, the minima of the PEL correspond to mechanically stable arrangement of particles \cite{Stillinger_Science_1995_IGS}. A correlation is drawn between the fragility of a glass former and the density of minima in the PEL. Fragile glass formers are observed to have more density of minima than strong glass formers \cite{Angell_J_Non_Cryst_Solids_1991_IGS}. Interestingly, the PEL is particularly useful to calculate another important quantity, the configurational entropy ($S_{c}$) of a supercooled liquid \cite{Sastry_Nature_2001_IGS}. In the supercooled regime, $S_{c}$ is the excess entropy of the liquid over its crystalline phase. At a hypothetical temperature known as Kauzmann temperature $T_{K}$, $S_{c}$ vanishes and the system is postulated to go through a thermodynamic transition to avoid a catastrophe (the Kauzmann catastrophe) which would require the entropy of the crystalline state to be greater than that of the liquid state. The resultant `ideal glass' state corresponds to the global minimum of the PEL \cite{Debenedetti_Stillinger_Nature_2001_IGS}.\\
\indent Even after decades of research, a proper understanding of the ideal glass state and the link between the thermodynamics and kinetics of glass formers remains elusive. However, it is seen that for most fragile supercooled liquids, $T_{K}$ is approximately equal to $T_{0}$, indicating a possible relation between its thermodynamics and kinetics \cite{Bohmer_Ngai_Angell_Plazek_JCP_1993_IGS}. The random first order transition theory of the glass transition predicts a possible relation between the stretching exponent $\beta$ of the non-Debye $\alpha$-relaxation and $T_{K}$ \cite{Xia_Wolynes_PRL_2001_IGS}. It is believed that $\beta$ is temperature dependent and vanishes at $T_{K}$ which corresponds to the divergence of the width $w$ of the $\alpha$-relaxation spectrum \cite{Xia_Wolynes_PRL_2001_IGS,Dixon_Nagel_PRL_1988_IGS}.\\
\indent In this work, we have carried out detailed experimental measurements of the microscopic relaxation timescales of colloidal suspensions of Laponite approaching the jamming transition by dynamic light scattering experiments. The evolution of microscopic relaxation timescales are studied for Laponite suspensions with different concentrations ($C_{L}$), added salt concentrations ($C_{S}$) and temperatures ($T$) to observe the effects of these variables on the fragility parameter $D$. It is seen that the evolutions of relaxation timescales of Laponite for all $C_{L}$, $C_{S}$ and $T$ can be compared with molecular glass formers if the waiting time ($t_{w}$) of the former is mapped with the inverse of thermodynamic temperature ($1/T$) of the latter \cite{Debasish_YMJ_Ranjini_Soft_Matter_2014_IGS}. The stretching exponent $\beta$, associated with non-Debye structural relaxation processes, is also extracted for different waiting times $t_{w}$ for samples of different $C_{L}$, $C_{S}$ and $T$. A hypothetical timescale $t_{k}$ at which the width of the distribution of structural relaxation times diverges is defined by extrapolating $\beta\rightarrow 0$. It is shown here that this timescale is correlated to the timescale $t^{\infty}_{\alpha}$ at which the mean structural relaxation time diverges. This remarkable correlation between these two hypothetical timescales is reminiscent of the correlation of two hypothetical temperatures, Kauzmann and Vogel temperatures, seen previously in extremely fragile molecular glass formers. We explain this correlation by appropriately modifying the coupling model for molecular glass formers and by analysing the observed waiting time dependence of the stretching exponent associated with the primary relaxation process of Laponite particles in suspension.
\section{Sample preparation and Experimental methods}
Laponite RD $\circledR$ (BYK Inc.) is dried in an oven for more than 16 hours to remove the absorbed water. A homogeneous and optically clear aqueous suspension is prepared by vigorous mixing of Laponite with Millipore water (resistivity 18.2 M$\Omega$-cm) and subsequent filtration at a constant flow rate (3.0 ml/min) by a syringe pump (Fusion 400, Chemyx Inc.). Salt concentration is adjusted by precise addition of a concentrated solution of NaCl (Sigma-Aldrich) of a known molarity by a pipette under vigorous mixing. Next, the Laponite suspension is sealed immediately in a cuvette. Aging time or waiting time $t_{w}$ is calculated from the moment the sample is sealed. DLS experiments are performed by a Brookhaven Instruments Corporation BI-200SM spectrometer equipped with a temperature controller (Polyscience Digital). Details of the set-up are given elsewhere \cite{Debasish_YMJ_Ranjini_Soft_Matter_2014_IGS}. The normalized intensity autocorrelation function of the scattered light, $g^{(2)}(t) = \frac{<I(0)I(t)>}{<I(0)>^{2}} =  1+ A|g^{(1)}(t)|^{2}$, is obtained as a function of delay time $t$ from a digital autocorrelator (Brookhaven BI-9000AT). Here, $I(t)$, $g^{(1)}(t)$ and $A$ are the intensity of the scattered light at a delay time $t$, the normalized electric field autocorrelation function and the coherence factor respectively \cite{Bern_Pecora_IGS}.
\section{Results and Discussions}
\begin{figure}
\includegraphics[width=4in]{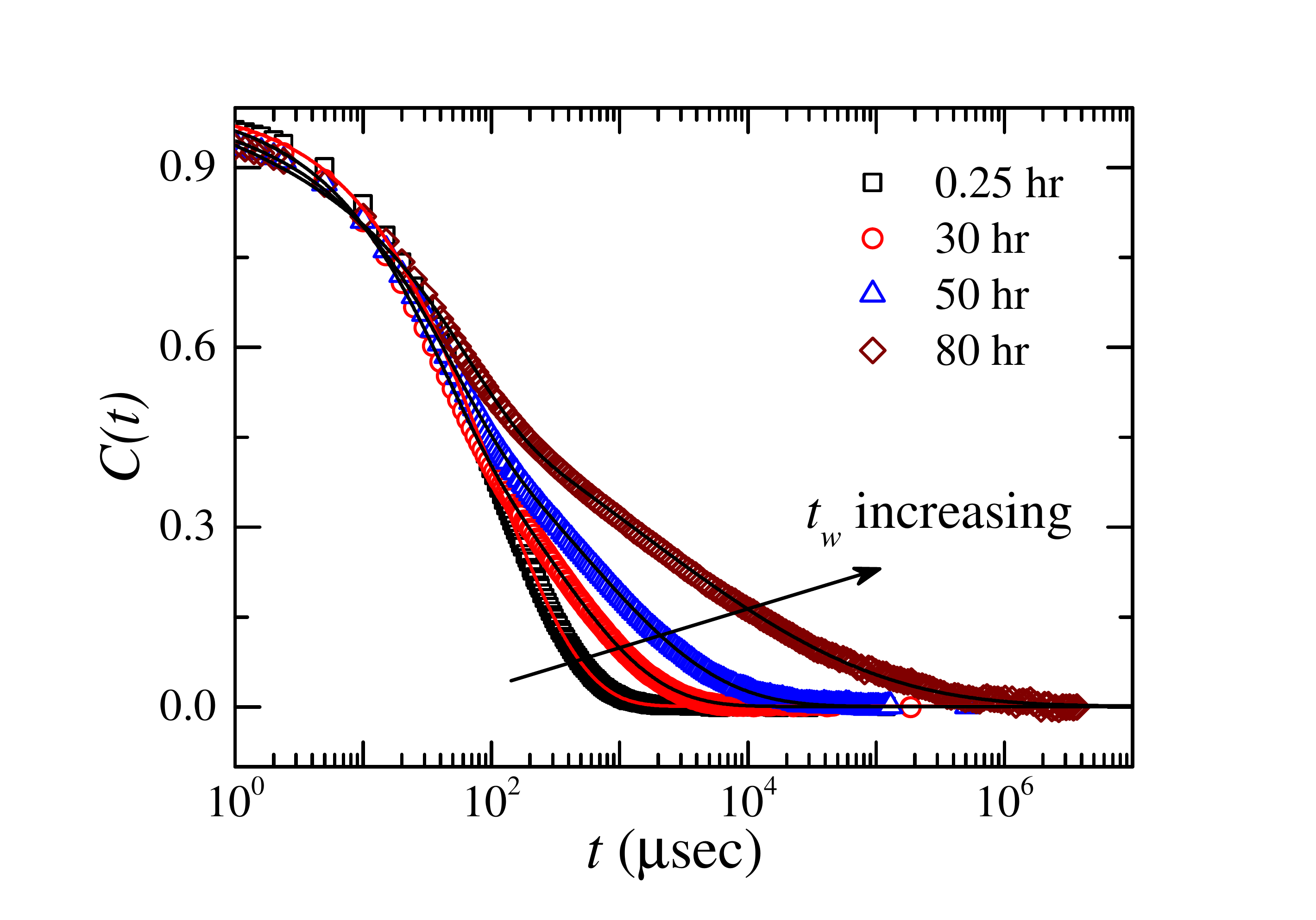}
\caption{The normalized intensity autocorrelation functions $C(t)$, {\it vs.} the delay time $t$, at 25$^\circ$C and scattering angle $\theta$ = 90$^\circ$, for 2.5\% w/v Laponite suspension with 0.05 mM salt at several $t_{w}$. The solid lines are fits to equation~\ref{autocorrelation_function_IGS}.}
\label{autocorrelation_IGS}
\end{figure}
\begin{figure}
\includegraphics[width=6in]{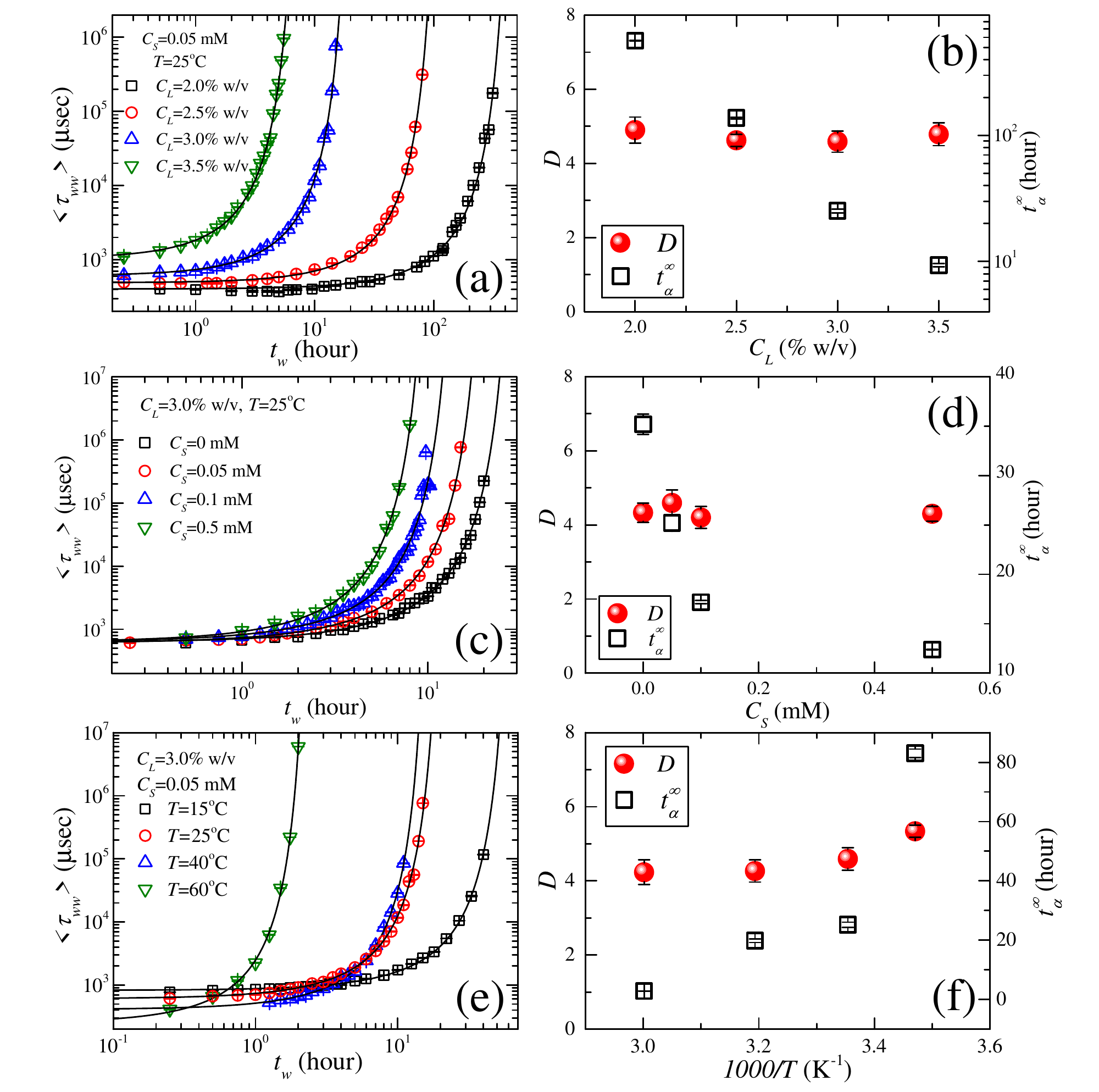}
\caption{Mean slow relaxation times $<\tau_{ww}>$, obtained by fitting $C(t)$ data to equation~\ref{autocorrelation_function_IGS}, are plotted {\it vs.} $t_{w}$ for different $C_{L}$, $C_{S}$ and $T$ in (a), (c) and (e) respectively. Fragility $D$ ($\bullet$) and Vogel time $t^{\infty}_{\alpha}$ ($\square$), measured by fitting $<\tau_{ww}>$ data to the equation~\ref{VFT_IGS}, are plotted for different $C_{L}$, $C_{S}$ and $T$ in (b), (d) and (f) respectively.} 
\label{mean_tau_D_vogel_time}
\end{figure}
\begin{figure}
\includegraphics[width=4in]{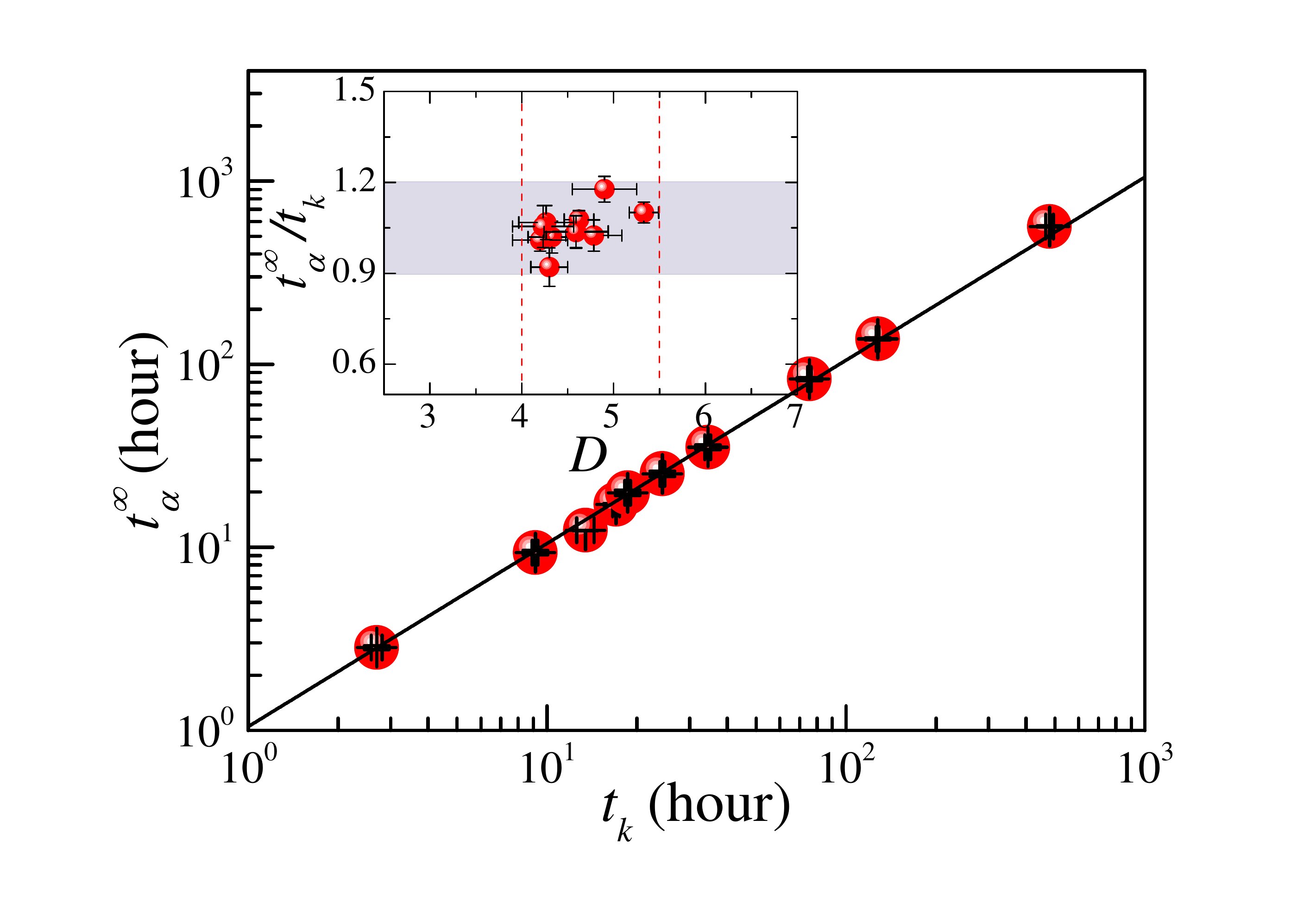}
\caption{Analogous Vogel time $t^{\infty}_{\alpha}$ is plotted {\it vs.} $t_{k}$, the waiting time at which $\beta\rightarrow 0$. Solid line is the linear fit with a slope 1.02$\pm$0.025. In the inset, the ratio $t^{\infty}_{\alpha}/t_{k}$ is plotted {\it vs.} $D$.}
\label{tk_tinf_D_correlation}
\end{figure}

Intensity autocorrelation functions $g^{(2)}(t)$ are obtained as a function of the waiting time $t_{w}$. In figure~\ref{autocorrelation_IGS}, the normalized intensity autocorrelation function, $C(t) = g^{(2)}(t)-1$, for a 2.5\% w/v Laponite suspension with 0.05 mM salt at $25^{\circ}$C, is plotted as a function of delay time $t$ for several $t_{w}$ values. Two-step relaxation processes, which become more prominent as $t_{w}$ increases, are observed in $C(t)$. It is also seen that the decay of $C(t)$ slows down with $t_{w}$ and can be expressed by a sum of an exponential and a stretched exponential decay in the following way \cite{Debasish_YMJ_Ranjini_Soft_Matter_2014_IGS,Ruzicka_JPCM_2004_IGS,Ruzicka_PRL_2004_IGS}.\\
\begin{equation}
\label{autocorrelation_function_IGS}
C(t)=[a\exp{\left\{-t/\tau_{1}\right\}}+(1-a)\exp{\left\{-(t/\tau_{ww})^{\beta}\right\}}]^{2}
\end{equation}
\indent The data in figure~\ref{autocorrelation_IGS} are fitted to equation~\ref{autocorrelation_function_IGS}, with $\tau_{1}$, $\tau_{ww}$, $a$ and $\beta$ being the fitting parameters. The fast decay is expressed by an exponential function and describes a relaxation time quantified by a timescale $\tau_{1}$, the fast relaxation time. Similarly, the slow decay is represented by a stretched exponential function, where $\tau_{ww}$ and $\beta$ are the slow relaxation time and stretching exponent respectively \cite{Ruzicka_JPCM_2004_IGS}. The mean slow relaxation time is given by $<\tau_{ww}>=(\tau_{ww}/\beta)\Gamma(1/\beta)$ \cite{Lindsey_JCP_1980_IGS}.\\
\indent Two-step decays are often seen in glass formers \cite{Angell_Ngai_McKenna_McMillan_Martin_JAP_2000_IGS,Mezei_Knaak_Farago_PRL_1987_IGS,Sidebottom_et_al_PRL_1993_IGS}. In supercooled liquids and in the present system, the faster decay involves diffusion of a particle within a cage formed by its neighbours \cite{Gotze_MCT_IGS}, while the slower decay ($\alpha$-relaxation process) is connected to structural or orientational rearrangements \cite{Bohmer_Ngai_Angell_Plazek_JCP_1993_IGS}. The mean slow relaxation time $<\tau>$ is very sensitive to changes in temperature and can be expressed by the Vogel-Fulcher-Tammann relation $<\tau>=\tau_{VF}\exp[B/(T-T_{0})]$ \cite{Bohmer_Ngai_Angell_Plazek_JCP_1993_IGS}, where the fitting parameter $\tau_{VF}=<\tau>(T\rightarrow\infty)$. $B$ and $T_{0}$ are identified as the fragility index and the Vogel temperature respectively. For quantifying the deviation from Arrhenius behavior, i.e. $<\tau>=\tau\exp(E/k_{B}T)$, one can define $B=DT_{0}$, where $D$ is the strength or fragility parameter \cite{Bohmer_Ngai_Angell_Plazek_JCP_1993_IGS,Angell_J_Non_Cryst_Solids_1991_IGS,Lubchenko_Wolynes_Annu_Rev_Phys_Chem_2007_IGS}. As discussed earlier in the introduction, the slowing down of the relaxation processes in Laponite suspension can be compared to the dynamics of supercooled liquids by performing a one-to-one mapping between the waiting time $t_{w}$ of a spontaneously evolving Laponite suspension and the inverse of the temperature $1/T$ of a supercooled liquid that is quenched towards the glass transition \cite{Debasish_YMJ_Ranjini_Soft_Matter_2014_IGS}. The mapping works well as the reduction in mobility with decrease in temperature in fragile molecular glass formers is analogous to the decrease in particulate mobility in Laponite suspensions with increase in waiting time. It was also reported that $\tau_{1}$ increases exponentially with $t_{w}$ i.e. $\tau_{1}=\tau^{0}_{1}\exp(t_{w}/t^{\infty}_{\beta})$, where $t^{\infty}_{\beta}$ is the characteristic time associated with the slowdown of the fast relaxation process. Simultaneously, $<\tau_{ww}>$ shows a VFT-type dependence on $t_{w}$ given by the following expression \cite{Debasish_YMJ_Ranjini_Soft_Matter_2014_IGS}:
\begin{equation}
<\tau_{ww}> = <\tau_{ww}>^{0}\exp(Dt_{w}/(t^{\infty}_{\alpha}-t_{w}))
\label{VFT_IGS}
\end{equation}
\indent In equation~\ref{VFT_IGS}, $D$ is the fragility parameter and $t^{\infty}_{\alpha}$ is identified with the Vogel time or the waiting time at which $<\tau_{ww}>$ diverges.\\
%
\indent In figure~\ref{mean_tau_D_vogel_time}(a), $<\tau_{ww}>$ values extracted from fits of the autocorrelation data to equation~\ref{autocorrelation_function_IGS} and the corresponding fits to equation~\ref{VFT_IGS}, are plotted for different values of $C_{L}$ with $C_{S}$=0.05 mM and $T$=25$^{\circ}$C. It is seen from the plots that the evolution of $<\tau_{ww}>$ becomes faster with increase in $C_{L}$ \cite{Debasish_et_al_Langmuir_2015_IGS}. The fragility parameters $D$ ($\bullet$) and the Vogel times $t^{\infty}_{\alpha}$ ($\square$) obtained from the fits are plotted in figure~\ref{mean_tau_D_vogel_time}(b). Recent simulation results suggest that the kinetic fragility $K_{VFT}$ ($K_{VFT}=1/D$), calculated from the $\alpha$-relaxation time for Kob-Anderson (KA) model glass formers has a very weak dependence on density \cite{Shiladitya_Sastry_EPJE_2013_IGS}. In the Laponite suspensions studied here, the suspension density changed from 1020 Kgm$^{-3}$ to 1035 Kgm$^{-3}$ when $C_{L}$ is varied between 2.0\% w/v and 3.5\% w/v. Since the change in the density of Laponite suspensions studied here is very small, $D$ can be expected to be constant. Remarkably, our experimental observation, demonstrated in figure~\ref{mean_tau_D_vogel_time}(b), supports this observation for molecular glass formers.\\
\indent It is to be noted that although the bulk suspension density does not change appreciably within the range of $C_{L}$ studied here, the number density of particles increases with $C_{L}$. The subsequent decrease in the interparticle distance in Laponite suspensions for higher $C_{L}$ therefore translates to an increase in pressure as more particles are now packed in the same volume of the suspension. The apparent independence of $D$ on number density in the Laponite suspensions studied here is therefore reminiscent of recent results for many molecular glass formers for which the isochoric fragility is independent of pressure \cite{Casalini_Roland_PRB_2005_IGS}. It is also seen from figure~\ref{mean_tau_D_vogel_time}(b) that $t^{\infty}_{\alpha}$ decreases monotonically with $C_{L}$, thereby indicating a faster approach to an arrested state with increasing Laponite concentration. Since the number density of Laponite particles in suspension is directly proportional to $C_{L}$, multi-body interactions are enhanced with increasing $C_{L}$, thereby shifting the onset of the glass transition to earlier times (figure~\ref{mean_tau_D_vogel_time}(a)). This feature is reminiscent of a previous observation in dense colloidal suspensions \cite{Stickel_Powell_Annu_Rev_Fluid_Mech_2005_IGS,Masri_Petekidis_Berthier_et_al_JStat_2009_IGS} and has been also discussed earlier in the context of the aging of Laponite suspensions \cite{Debasish_YMJ_Ranjini_Soft_Matter_2014_IGS,Debasish_et_al_Langmuir_2015_IGS}.\\
\indent In figure~\ref{mean_tau_D_vogel_time}(c), $<\tau_{ww}>$ is plotted for different values of $C_{S}$ with $C_{L}$=3.0\% w/v and $T$=25$^{\circ}$C \cite{Debasish_et_al_Langmuir_2015_IGS}. $D$ ($\bullet$) and $t^{\infty}_{\alpha}$ ($\square$) are obtained from the fits of the data to equation~\ref{VFT_IGS} and are plotted in figure~\ref{mean_tau_D_vogel_time}(d). $D$ is almost constant for the entire range of salt concentrations $C_{S}$. DLVO calculations for Laponite suspensions reported earlier have revealed that the height of the repulsive barrier increases and the width of the barrier decreases with increase in $C_{S}$ \cite{Debasish_et_al_Langmuir_2015_IGS}. This is due to the enhancement of the screening of the interparticle repulsive interactions and the increasingly important role that interparticle attractions play in spontaneously evolving Laponite suspensions. The addition of salt and the development of interparticle attractions clearly increase the rate of structure formation. This is verified in figure~\ref{mean_tau_D_vogel_time}(d), where $t^{\infty}_{\alpha}$ is seen to decrease rapidly with increase in $C_{S}$ \cite{Debasish_et_al_Langmuir_2015_IGS}. This corroborates our earlier observation in \cite{Debasish_et_al_Langmuir_2015_IGS} that the arrested state is approached at a faster rate due to stronger interparticle interactions when the concentration of salt in the system is enhanced. However, an almost constant value of $D$ (figure~\ref{mean_tau_D_vogel_time}(d)) with increasing $C_{S}$ indicates that the fragility parameter is independent of the screening effects on the interparticle interaction within the range of salt concentrations studied here.\\
\indent Simulation results for binary mixtures of soft spheres have shown that fragility is independent of the softness of the repulsive interaction \cite{Michele_Sciortino_Coniglio_JPCM_2004_IGS}. However, recent computer simulation results on binary mixture glass formers with a modified Lennard-Jones type potential show that the kinetic fragility increases with increasing softness \cite{Sengupta_Sastry_JCM_2011_IGS}. Experiments on soft colloidal systems show that soft (more compressible) particles form stronger glasses than hard (less compressible) colloidal particles \cite{Mattsson_Miyazaki_Nature_2009_IGS}. In our experiment, it is to be noted that while the proportion of attractive and repulsive interaction changes with $C_{S}$, the softness of interparticle interactions and the compressibility of the Laponite particles do not change. This is established from the observed self-similarity of the potential energy landscape with $C_{S}$ reported earlier \cite{Debasish_et_al_Langmuir_2015_IGS}. Hence, the apparent insensitivity of $D$ to changes in $C_{S}$ confirms several earlier simulations and experimental results on colloidal and molecular glass formers \cite{Michele_Sciortino_Coniglio_JPCM_2004_IGS,Sengupta_Sastry_JCM_2011_IGS,Mattsson_Miyazaki_Nature_2009_IGS}.\\
%
%
\indent Temperature has a strong effect on the evolution of relaxation processes in Laponite suspensions \cite{Debasish_et_al_Langmuir_2015_IGS,Shahin_Joshi_Langmuir_2012_IGS}. $<\tau_{ww}>$ is plotted for different values of $T$ with $C_{L}$=3.0\% w/v and $C_{S}$=0.05 mM in figure~\ref{mean_tau_D_vogel_time}(e). It is seen that increase in $T$ accelerates the time-evolution of $<\tau_{ww}>$ \cite{Debasish_et_al_Langmuir_2015_IGS,Shahin_Joshi_Langmuir_2012_IGS}. In figure~\ref{mean_tau_D_vogel_time}(f), $D$ ($\bullet$) and $t^{\infty}_{\alpha}$ ($\square$) are plotted for the data shown in figure~\ref{mean_tau_D_vogel_time}(e). It is seen that $D$ is weakly dependent on temperature. $D$ increases by a small amount as $T$ decreases. However, $t^{\infty}_{\alpha}$ is seen to increase with decrease in temperature, thereby indicating that the glass transition is achieved at earlier times with increase in $T$. Increase in $T$ is therefore equivalent to increasing the apparent cooling rate $q'$, with the system being driven towards the glass transition at faster rates at higher temperatures. For bulk metallic glass formers, i.e. Zr$_{57}$Cu$_{15.4}$Ni$_{12.6}$Al$_{10}$Nb$_{5}$ and Zr$_{58.5}$Cu$_{15.6}$Ni$_{12.8}$Al$_{10.3}$Nb$_{2.8}$, there is an apparent increase in $D$ at slower cooling rates \cite{Evenson_Gallino_Busch_JAP_2010_IGS}, while the Vogel temperature $T_{0}$ increases with cooling rate. The small increase in $D$ with $1000/T$ at the slowest cooling rate seen in figure~\ref{mean_tau_D_vogel_time}(f) therefore, is in accordance with the observations in \cite{Evenson_Gallino_Busch_JAP_2010_IGS}. Since an inverse relation exists between $T_{0}$ and $t^{\infty}_{\alpha}$ (i.e. $T_{0}\leftrightarrow 1/t^{\infty}_{\alpha}$) \cite{Debasish_YMJ_Ranjini_Soft_Matter_2014_IGS}, the decrease of $t^{\infty}_{\alpha}$ with increase in $T$ or $q'$ is remarkably consistent with the observations in metallic glass formers \cite{Evenson_Gallino_Busch_JAP_2010_IGS}.\\
\indent It is seen that strong glass formers have a lower density of minima in the potential energy landscape compared to fragile glass formers \cite{Angell_J_Non_Cryst_Solids_1991_IGS}. At present, the very small changes in $D$ reported in figure~\ref{mean_tau_D_vogel_time} indicate that the underlying energy landscapes are self-similar for the ranges of $C_{L}$, $C_{S}$ and $T$ studied here. This is in agreement with the conclusion from very recent experimental observations that time-evolutions of the microscopic relaxation timescales and the stretching exponents $\beta$ associated with the slow relaxation process show comprehensive Laponite concentration-salt concentration-temperature-waiting time superpositions, thereby indicating the self-similarity of the energy landscapes \cite{Debasish_et_al_Langmuir_2015_IGS}.\\ 
\begin{figure}
\includegraphics[height=6.8in]{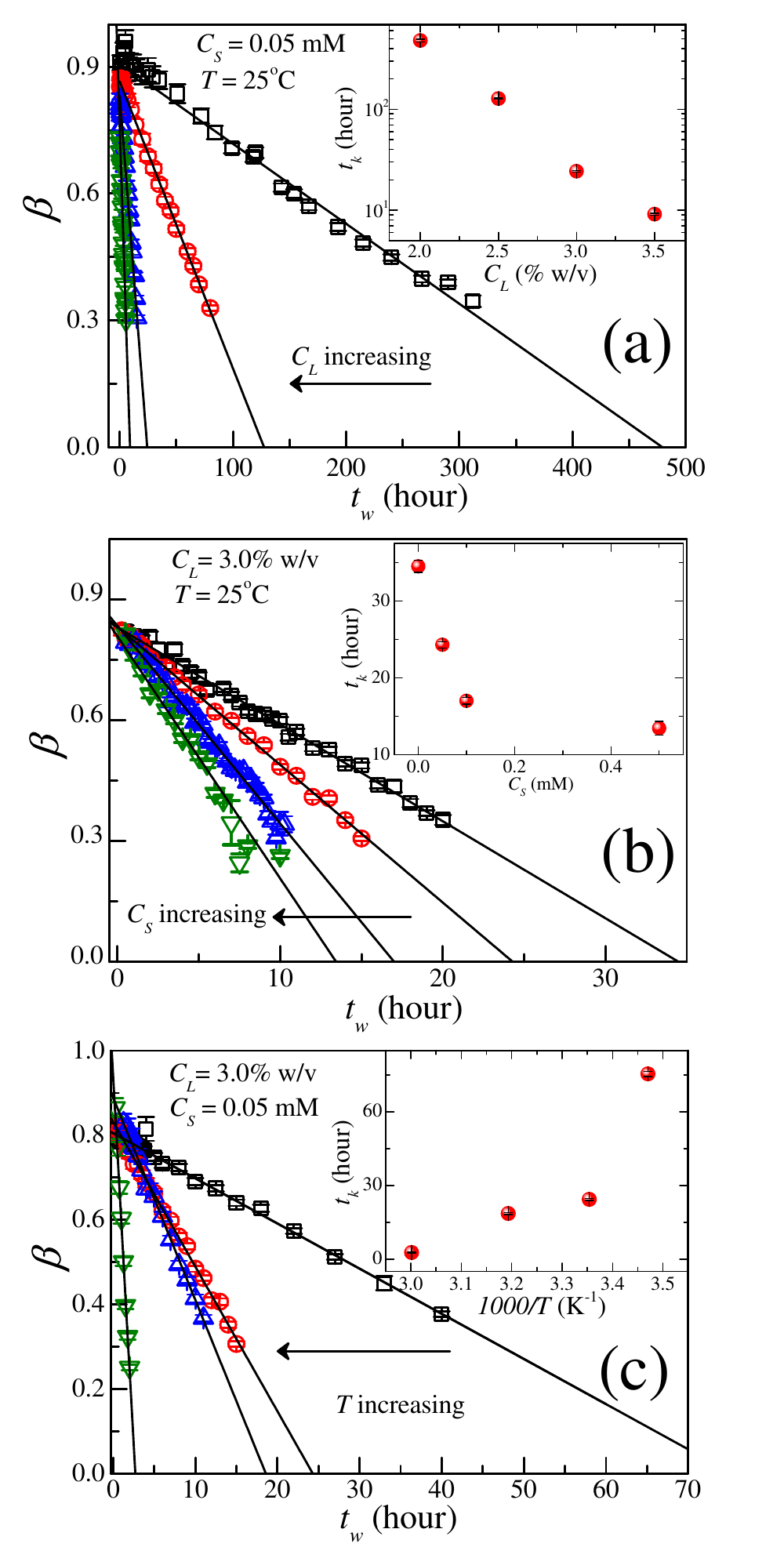}
\caption{(a) Stretching exponents $\beta$ are plotted {\it vs.} $t_{w}$ for $C_{S}$=0.05 mM and $T$=25$^{\circ}$C at different $C_{L}$ (from right to left, 2.0\% w/v ($\square$), 2.5\% w/v ($\circ$), 3.0\% w/v ($\triangle$), 3.5\% w/v ($\nabla$)). (b) $\beta$ {\it vs.} $t_{w}$ for $C_{L}$=3.0\% w/v and $T$=25$^{\circ}$C at different $C_{S}$ (from right to left, 0 mM ($\square$), 0.05 mM ($\circ$), 0.1 mM ($\triangle$), 0.5 mM ($\nabla$)). (c) $\beta$ {\it vs.} $t_{w}$ for $C_{L}$=3.0\% w/v and $C_{S}$=0.05 mM at different $T$ (from right to left, 15$^{\circ}$C ($\square$), 25$^{\circ}$C ($\circ$), 40$^{\circ}$C ($\triangle$), 60$^{\circ}$C ($\nabla$)). $t_{k}$ ($\bullet$), the waiting times at which $\beta\rightarrow 0$, measured by extrapolation of $\beta$ to 0, are plotted for different $C_{L}$, $C_{S}$ and $T$ in the insets of (a), (b) and (c) respectively.}
\label{Beta_tk}
\end{figure}
\indent In figure~\ref{Beta_tk}, the stretching exponent $\beta$, obtained by fitting the $C(t)$ data to equation~\ref{autocorrelation_function_IGS}, is plotted {\it vs.} $t_{w}$ for different $C_{L}$, $C_{S}$ and $T$. It is seen that for all $C_{L}$, $C_{S}$ and $T$, $\beta$ decreases linearly with $t_{w}$. For many supercooled liquids, $\beta$ depends on temperature and decreases linearly with $1/T$ \cite{Dixon_Nagel_PRL_1988_IGS}. As discussed earlier, a non-Arrhenius stretched exponential decay of the $C(t)$ data can arise due to the well-known Kohlrausch-Williams-Watts distribution of the slow relaxation timescales given by $\rho_{ww}(\tau)=-\frac{\tau_{ww}}{\pi\tau^2}\sum^{\infty}_{k=0}\frac{(-1)^k}{k!}\sin(\pi\beta k)\Gamma(\beta k+1)\left(\frac{\tau}{\tau_{ww}}\right)^{\beta k+1}$ \cite{Lindsey_JCP_1980_IGS}. The width $w$ of the distribution can be written as $w=\frac{<\tau^{2}_{ww}>}{<\tau_{ww}>^{2}}-1=\frac{\beta\Gamma(2/\beta)}{(\Gamma(1/\beta))^{2}}-1$. Here $\beta\leq 1$, with a lower value of $\beta$ indicating a broader distribution of relaxation timescales, with the width $w$ of the distribution $\rho_{ww}(\tau)$ diverging at $\beta\rightarrow 0$ \cite{Lindsey_JCP_1980_IGS}. We define a time $t_{k}$ as the waiting time at which $\beta\rightarrow 0$. This definition of a hypothetical divergence time $t_{k}$ is similar to the definition of the Kauzmann temperature $T_{K}$ for molecular glass formers where it is seen that $w$ diverges with a vanishing $\beta$ at $T_{K}$ \cite{Xia_Wolynes_PRL_2001_IGS,Dixon_Nagel_PRL_1988_IGS,Dixon_PRB_1990_IGS,Papoular_Phil_Mag_Lett_1991_IGS}. From the data plotted in figure~\ref{Beta_tk}, $t_{k}$ is measured by extrapolating $\beta$ to 0 for different $C_{L}$, $C_{S}$ and $T$. It is seen from the insets of figures~\ref{Beta_tk}(a)-(c) that the extrapolated values of $t_{k}$ decrease with the increase of $C_{L}$, $C_{S}$ and $T$.\\
%
\indent The temperature $T_{K}$ corresponding to a vanishing $\beta$ in fragile supercooled liquids \cite{Dixon_PRB_1990_IGS,Papoular_Phil_Mag_Lett_1991_IGS}, which is typically calculated from the extrapolation of the temperature dependent $\beta$ data, is seen to be correlated to $T_{0}$ \cite{Dixon_Nagel_PRL_1988_IGS,Xia_Wolynes_PRL_2001_IGS}, i.e. $T_{0}\approx T_{K}$. Given the many similarities between supercooled liquids and aging Laponite suspensions, it is interesting to investigate if $t_{k}$ values calculated for the latter have any connection with $t^{\infty}_{\alpha}$, the analogous Vogel time. In figure~\ref{tk_tinf_D_correlation}, we plot $t^{\infty}_{\alpha}$ {\it vs.} $t_{k}$. Remarkably, it is seen that $t_{k}$ is correlated with $t^{\infty}_{\alpha}$, {\it i.e.} $t^{\infty}_{\alpha}\approx t_{k}$. We further note that for all Laponite suspensions studied here, $4\leq D\leq5.5$ (inset of figure~\ref{tk_tinf_D_correlation}). It is to be noted that values of $D$ for sorbitol, toluene, {\it o}-terphenyl, propylene carbonate, triphenyl phosphite and sucrose are 8.6, 5.6, 5.0, 2.9, 2.9 and 0.154 respectively \cite{Tanaka_PRL_2003_relationship_Thermodynamics_Kinetics_IGS,Angell_J_Res_Natl_Inst_Stand_Technol_1997_IGS}. The obtained values of $D$ for Laponite suspensions therefore compare very well with the $D$-values for these fragile molecular glass formers, thereby indicating that Laponite suspensions are excellent candidates for the study of the glass transition \cite{Kivelson_Tarjus_Nature_Materials_Commentary_2008_IGS}.\\
\indent In the supercooled liquids literature, the Kauzmann temperature $T_{K}$ is also known as the equilibrium glass transition temperature \cite{Adam_Gibbs_JCP_1965_IGS}. According to this theory, the system in the supercooled regime relaxes by exploring possible configurations available in the energy landscape via activated processes. A good measure of the number of available configurations at a particular temperature can be obtained by calculating the configurational entropy at that temperature and is given by $S_{c}(T)=S_{liquid}(T)-S_{crystal}(T)$ when $S_{liquid}$ is the entropy of the liquid and $S_{crystal}$ is the entropy of the crystalline state, with $S_{c}(T)$ vanishing as $T\rightarrow T_{K}$. However, it is not completely clear that a simple relation between the kinetics and the thermodynamics of a glass former ({\it i.e.} $T_{0}\approx T_{K}$) always exists, as a systematic increase of $T_{k}/T_{0}$ from unity was observed for increasing values of $D$ \cite{Tanaka_PRL_2003_relationship_Thermodynamics_Kinetics_IGS}. It was reported in an earlier work that the ratio $T_{k}/T_{0}$ lies between 0.9 to 1.2 for $D<20$ \cite{Tanaka_PRL_2003_relationship_Thermodynamics_Kinetics_IGS}. We next plot the ratio $t^{\infty}_{\alpha}/t_{k}$ {\it vs.} $D$ for data acquired for all the Laponite suspensions of different $C_{L}$, $C_{S}$ and $T$ studied here in the inset of figure~\ref{tk_tinf_D_correlation}. It is seen that $0.9\leq t^{\infty}_{\alpha}/t_{k}\leq1.2$ for all the Laponite suspensions. Indeed, it was shown in an earlier study that $T_{K}/T_{0}$, the ratio of the Kauzmann and Vogel temperatures, starts increasing from unity for fragile glass formers and attains higher values for strong supercooled liquids \cite{Tanaka_PRL_2003_relationship_Thermodynamics_Kinetics_IGS}. Although there is an apparent similarity between the results plotted in figure~\ref{tk_tinf_D_correlation} and $T_{k}/T_{0}$ ratios reported earlier \cite{Tanaka_PRL_2003_relationship_Thermodynamics_Kinetics_IGS}, we note that $t_{k}$ is the time at which $w\rightarrow\infty$. Clearly, for the colloidal suspensions of Laponite studied here, $t_{k}$ has a very different physical origin when compared to $T_{K}$ in supercooled liquids.\\
%
\indent We now explain the simultaneous divergence of the two hypothetical times, $t^{\infty}_{\alpha}$ and $t_{k}$ in terms of the Vogel-Fulcher behavior of the relaxation times and the stretched-exponential nature of the slow relaxation process \cite{Nagel_Dixon_JCP_1989_IGS}. The observed stretched exponential time dependence of the slow relaxation process can be rewritten for aging colloidal suspensions as:
\begin{equation}
\exp\left\{-\left(t/\tau\right)^{\beta(t_{w})}\right\}=\exp\left\{-t/L(t_{w},t)\right\}
\label{eq:L}
\end{equation} 
Here, $L(t_{w},t)$ can be interpreted as the time-dependent relaxation time and has a similar physical origin as for supercooled liquids \cite{Nagel_Dixon_JCP_1989_IGS}. Following the coupling model proposed by Ngai {\it et. al.} \cite{Ngai_Rendell_Rajagopal_Teitler_Ann_New_York_Acad_Sci_1986_IGS,Ngai_Rajagopal_Teitler_JCP_1988_IGS,Rajagopal_Teitler_Ngai_JPC_1984_IGS}, we relate $L(t_{w},t)$ to the time-dependent relaxation rate $W(t)$ by $W(t)=\beta/L(t_{w},t)$ \cite{Ngai_et_al_comment_JCP_1989_IGS}. This model of relaxation indicates that each relaxation unit of the system relaxes independently with a primitive rate $W_{0}$ at timescales that are short when compared to the characteristic timescale $t_{0}=1/\omega_{c}$ associated with the coupling of the relaxing molecular units. At longer times, i.e. $\omega_{c}t>1$, the primitive relaxation rate slows down and can be expressed as, $W(t)=W_{0}(\omega_{c}t)^{-n}$, 0$<n<$1 and $n=1-\beta$ \cite{Ngai_et_al_comment_JCP_1989_IGS}. If $L(t_{w},t)=L_{0}(t_{w})(t/t_{0})^{1-\beta}$ \cite{Nagel_Dixon_JCP_1989_IGS}, where $L_{0}=\beta/W_{0}$ \cite{Ngai_et_al_comment_JCP_1989_IGS}, it follows from equation~\ref{eq:L} that,
\begin{equation}
\exp\left[-\left(t/\tau\right)^{\beta}\right]=\exp\left[-\frac{t}{L_{0}(t_{w})(t/t_{0})^{1-\beta}}\right]=\exp\left[-\left(\frac{t}{t_{0}[L_{0}(t_{w})/t_{0}]^{1/\beta}}\right)^{\beta}\right]
\end{equation}
Hence, $\tau$ can be rewritten as, $\tau=t_{0}[L_{0}(t_{w})/t_{0}]^{1/\beta}$ \cite{Nagel_Dixon_JCP_1989_IGS}. For colloidal suspensions of Laponite, the secondary relaxation process is related to the microscopic motion of a single relaxation unit (a Laponite particle) and has the following Arrhenius dependence on $t_{w}$ as discussed earlier \cite{Debasish_YMJ_Ranjini_Soft_Matter_2014_IGS}: $\tau_{1}=\tau^{0}_{1}\exp\left(t_{w}/t^{\infty}_{\beta}\right)$. As $L_{0}(t_{w})$ is related to the single relaxation unit \cite{Ngai_et_al_comment_JCP_1989_IGS}, we assume an Arrhenius dependence of $L_{0}(t_{w})$ on $t_{w}$ for colloidal suspension of Laponite, $L_{0}(t_{w})=L\exp\left(t_{w}/t^{\infty}_{\beta}\right)$, where $L$ is a constant. We have already seen from figure~\ref{Beta_tk} that $\beta$ decreases linearly with $t_{w}$, i.e. $\beta(t_{w})=\beta_{0}(t_{k}-t_{w})$. Substituting the $t_{w}$-dependence of $\beta$ along with the Arrhenius dependence of $L_{0}(t_{w})$ in the expression for $\tau$ yields a VFT equation, 
\begin{equation}
\tau=t_{0}\left[\frac{L\exp\left(t_{w}/t^{\infty}_{\beta}\right)}{t_{0}}\right]^{1/\beta}=L^{1/\beta}t_{0}^{(\beta-1)/\beta}\exp\left(t_{w}/\beta t^{\infty}_{\beta}\right)=C\exp\left(\frac{t_{w}}{t^{\infty}_{\beta}\beta_{0}(t_{k}-t_{w})}\right)
\label{tau_coupling__tk_tinf_IGS}
\end{equation}
where $C=L^{1/\beta}t_{0}^{(\beta-1)/\beta}$. Equation~\ref{tau_coupling__tk_tinf_IGS} has an identical form as equation~\ref{VFT_IGS}, when $t_{k}\approx t^{\infty}_{\alpha}$, with $\tau$ diverging at $t_{w}\rightarrow t^{\infty}_{\alpha}$ and the width $w$ of the distribution $\rho_{ww}(\tau)$ diverging simultaneously at $t_{k}$. This explains the correlation between $t_{k}$ and $t^{\infty}_{\alpha}$ observed in figure~\ref{tk_tinf_D_correlation}.\\
\indent This successful adaptation of the coupling model, which was previously proposed for molecular glass formers, to the present scenario of aging colloidal Laponite suspensions, clearly demonstrates the universal nature of the approach to a final arrested state in these two seemingly different glass formers. Additionally, the comparable behavior of fragility parameter $D$ in these two glass formers and the explanation of the observed correlation between the two hypothetical divergence times $t^{\infty}_{\alpha}$ and $t_{k}$ using the coupling model together indicate a remarkable similarity of their relaxation processes at the particle scales.
\section{Conclusions}
In this work, the time evolutions of the relaxation processes of colloidal suspensions of Laponite are studied by dynamic light scattering (DLS). The fragility parameter $D$ is obtained by fitting the autocorrelation data $C(t)$ for Laponite suspensions of different concentrations ($C_{L}$), added salt concentrations ($C_{S}$) and temperatures ($T$). It is seen that the value of $D$ is approximately constant for the entire range of Laponite concentrations and salt concentrations investigated here. These results are reminiscent of the observed independence of the isochoric fragilities of supercooled liquids on pressure. Furthermore, $D$ is independent of the screening effects of the repulsive interparticle interactions due to the addition of salt. Finally, it is seen that $T$ determines the rate at which system approaches the glass transition (or the apparent cooling rate) and that $D$ is weakly dependent on $T$. This result is reminiscent of the dependence of the kinetic fragility on the cooling rate for metallic glass formers \cite{Evenson_Gallino_Busch_JAP_2010_IGS}.\\
\indent The stretching exponent $\beta$ for Laponite suspensions with different $C_{L}$, $C_{S}$ and $T$ is seen to decrease linearly with waiting time, indicating a divergence of the width of the relaxation time distribution at even higher waiting times. This observation is similar to the decrease of $\beta$ with $1/T$ in many fragile molecular glass formers. We next define a timescale $t_{k}$ at which the width of the distribution of the slow relaxation timescale diverges. We report a correlation between $t_{k}$ and $t^{\infty}_{\alpha}$, where $t^{\infty}_{\alpha}$ is the hypothetical Vogel time at which the average slow relaxation time diverges. This correlation corroborates analogous observations in fragile molecular glass formers for which it was reported that the Kauzmann and Vogel temperature are approximately equal ($T_{K}\approx T_{0}$). We next calculate the ratio $t^{\infty}_{\alpha}/t_{k}$ and plot it {\it vs.} $D$. This ratio is found be approximately 1 for all the $D$ values reported here. This observation is reminiscent of the change in the ratio $T_{K}/T_{0}$ with fragility parameter $D$ seen for several supercooled liquids \cite{Tanaka_PRL_2003_relationship_Thermodynamics_Kinetics_IGS}.\\
\indent Our results therefore clearly agree very well with existing results for fragile glass formers. Interestingly, in the case of Laponite which is a colloidal system, $t_{k}$ is measured from the kinetics of the relaxation process and signifies the time at which the width of the distribution of structural relaxation times diverges. In contrast, $T_{K}$ for supercooled liquids is a thermodynamic quantity and generally calculated from calorimetric data. The correlation between the two hypothetical diverging time scales for Laponite suspensions, $t_{k}\approx t^{\infty}_{\alpha}$, demonstrates that the average value and the width of the distribution of slow relaxation times diverge simultaneously. We explain this result using the coupling model proposed for molecular glass formers \cite{Ngai_Rendell_Rajagopal_Teitler_Ann_New_York_Acad_Sci_1986_IGS} and the $t_{w}$-dependence of the stretching exponent $\beta$ observed in our experiments.
%
%

%
%
\end{document}